\documentstyle[12pt]{article}
\normalbaselines
\begin{document}
\bibliographystyle{unsrt}

\def\IR{\relax{\rm I\kern-.18em R}}

\vbox {\vspace{6mm}}

\begin{center}
{\large \bf LINEAR TIME-DEPENDENT INVARIANTS FOR \\[2 mm]
SCALAR FIELDS AND NOETHER'S THEOREM}\footnote{Work
supported in part by project UNAM-DGAPA IN103091. }\\[7mm]
{\it O. Casta\~nos, R. L\'opez-Pe\~na and V.I. Man'ko}\footnote
{On leave from Lebedev Physics Institute, Moscow, Russia.} \\
{\it Instituto de Ciencias Nucleares, UNAM } \\
{\it Apdo. Postal 70-543, 04510 M\'exico, D. F., M\'exico} \\[5mm]

\end{center}

\vspace{2mm}

\begin{abstract}
The infinite number of time-dependent linear in field and conjugated
momenta invariants is derived for the scalar field using the Noether's
theorem procedure.
\end{abstract}

\vspace{5mm}

\noindent
PACS number(s): 03.65.Fd, 11.10.Ef.

\section{Introduction}

There exist in explicit form $2n$ integrals of motion linear in
position and momenta for $~n$-dimensional quantum systems, and its
classical counterpart, with Hamiltonians which are generic quadratic
forms in the position and momenta operators both with
time-independent and time-dependent coefficients \cite{mmt},
\cite{dod89}. Recently it was shown \cite{crm94} that these integrals
of motion may be obtained using standard Noether's theorem procedure.
It should be noted that for one-mode parametric oscillator
exists the Ermakov time-dependent invariant \cite{erm80} which
is quadratic in position and momentum.  The
variational derivation of this invariant using Noether's theorem has
been given in \cite{soliani}, \cite{santander1}, \cite{san2}, and
\cite{cfr91}.  The Noether's theorem derivation for the generalized $
2-D $ oscillator has been given in \cite{cr92}.  The quadratic
invariants in position and momenta using nonnoetherian symmetries
have been considered in \cite{hoj}.

On the other hand the linear
time-dependent integrals of motion for the quadratic field systems
have not been studied from the viewpoint of the Noether's theorem
procedure. The symplectic group $Sp (2\infty , \, \IR )$, producing
linear integrals of motion in the fields and their conjugated momenta
for a scalar field in two dimensions, has been shortly discussed in
Ref. \cite{dom}.

The aim of this work is to construct integrals of motion which are
linear forms in field and conjugated momentum to clarify their
noetherian symmetry nature for the scalar field obeying to the
Klein-Gordon equation. We will extend the approach given in
\cite{crm94} for a system with finite number of degrees of freedom
to the case of the field theory.

\section{Scalar field}
Let us consider neutral scalar field $\bar{\varphi} ( \vec{x} , \, t)
$ with mass $~m$ in four dimensions obeying the Klein-Gordon equation
	\begin{equation}
		(\frac {{\partial }^{2}}{{\partial t}^{2}} - \Delta +
		m^{2} ) \bar{\varphi} ( \vec{x} , \, t )= 0 \qquad
		(~c = 1 ).
		\end{equation}
For convenience we will study this field using periodical conditions
in the space
	\begin{equation}
		\bar{\varphi} ~( \vec{x} + \vec{L} ,~t) =
		\bar{\varphi} ~(\vec{x},~t),
		\end{equation}
i.e., embedding our system into a cube with characteristic length
$~L$.  The case of the infinite space is obtained by taking
appropriate limit $~L \rightarrow ~\infty $ \cite{ll}. Then by means
of the Fourier transform, the Eq.(1) can be reduced to an infinite
system of decoupled linear equations each describing the harmonic
oscillator
	\begin{equation}
		\bar{\varphi} ~(\vec{x} ,~t) = \sum _{\vec{k}}
		\varphi ~( \vec{k} ,~t)
		\exp (i\vec{k} \cdot \vec{x} ),
		\end{equation}
where the wave vector $ \vec{k} = 2 \pi \vec{n} / L $ and $ \vec{n} =
( n_{1} , \, n_{2} , \, n_{3} ) $, with $ n_{i} $ integer numbers.
The inverse Fourier transform gives the Fourier amplitudes $~\varphi
( \vec{k} ,\, t) $ in the form
	\begin{equation}
		\varphi \, (\vec{k},\, t) = \frac {1}{(2\pi L)^{3}}
		\int_{0}^{L} \int _{0}^{L}\int _{0}^{L} \bar{\varphi}
		\, (\vec{x},\, t) \exp (-i\vec{k} \cdot \vec{x})
		d\vec{x} \ .
		\end{equation}
Inserting the relation (3) into the Klein-Gordon Eq. (1) we obtain the
system of differential equations for each wave vector $\vec{k}$
	\begin{equation}
		\frac {\partial ^{2}\varphi \, (\vec{k},\, t)}{\partial
		t^{2}} + \omega ^{2} (\vec{k}) \varphi \, (\vec{k},\, t)
		= 0 \ ,
		\end{equation}
where the frequency is dependent on  field mode and mass $\, m$
	\begin{equation}
		\omega ^{2} (\vec{k}) = \vec{k}^{2} + m^{2},
		\end{equation}
and it is time-independent.

Thus we have in fact an infinite system of noninteracting
one-dimensional harmonic oscillators with coordinates $\, \varphi
\, (\vec{k},\, t)$.  The Lagrangian
as well as the Hamiltonian are infinite sums of the Lagrangians or
Hamiltonians of standard harmonic oscillators describing each field
mode, i.e.,
	\begin{equation}
		L = \sum_{ \vec{k} } ( \frac {\dot \varphi
		^{2}\, (\vec{k},\, t)}{2} - \omega^2 (\vec{k}) \ \frac
		{\varphi ^{2}\, (\vec{k},\, t)}{2} ) \ ,
		\end{equation}
and
	\begin{equation}
		H = \sum _{ \vec{k} } (\frac {\Pi ^{2}\, (\vec{k},\, t)}
		{2} +  \omega^2 (\vec{k}) \ \frac {\varphi^{2} (
		\vec{k} , \, t) }{2} ) \ ,
		\end{equation}
where the conjugated field momentum $\, \Pi\, (\vec{k},\, t)$ is related to
the field coordinate $\, \varphi \, (\vec{k},\, t)$ by the formula
	\begin{equation}
		\Pi\, (\vec{k},\, t) = \dot \varphi \, (\vec{k},\, t).
		\end{equation}
The classical systems with the Hamiltonian (8) have the
time-dependent invariants
	\begin{eqnarray}
		A_{\vec{k}}\, (t)&=&\exp (i\omega (\vec{k}) \, t) \frac
		{ \omega (\vec{k}) \ \varphi (\vec{k},t) + i \Pi
		(\vec{k},t) } {\sqrt {2  \omega (\vec{k})}}, \nonumber \\
		A_{\vec{k}}^{*}\, (t)&=&\exp (-i\omega (\vec{k}) \,
		t ) \frac { \omega (\vec{k}) \ \varphi (\vec{k},t) -
		i \Pi ( \vec{k}, \, t )}{\sqrt{2 \omega ( \vec{k} ) } }
		\ ,
		\end{eqnarray}
with the Poisson brackets
	\begin{equation}
		\{ A_{ \vec{k} } \, , \, A^{\ast}_ {\vec{k}^\prime }\} =
		-i\delta _{{\vec k \, , \, \vec k}^\prime } \ .
		\end{equation}
The quantization of the system is fulfilled in standard manner by the
replacement of classical variables $\, \varphi \, (\vec{k},\, t)$ and $\, \Pi
\, (\vec{k},\, t)$ by the operators with the canonical commutation
relations
	\begin{equation}
		[\hat \Pi _{\vec{k}} (\, t),\, \hat \varphi_{
		\vec{k}^\prime }(\, t)] = -i \delta _{{\vec k \, , \,
		\vec k}^\prime } \ , \qquad  \qquad \hbar = 1 \ .
		\end{equation}
Then the commutation relations of the quantized variables (Eq.(10))
are
	\begin{equation}
		[\hat A _{\vec{k}} (t),\,\hat A^{\dagger} _{\vec{k}^
		\prime }(t)] = \delta _{{\vec k \, , \, \vec
		k}^\prime } \ .
		\end{equation}
So the quantum scalar field has the infinite number of linear
invariants
	\begin{equation}
		\hat A _{\vec{k}} (t) = \exp (i\omega (\vec{k}) \,
		t) \frac {  \omega (\vec{k}) \ \hat \varphi
		(\vec{k},t) + i\hat \Pi (\vec{k},t)}{\sqrt {2 \omega
		( \vec{k} ) } }
		\end{equation}
with the commutation relations of boson creation and annihilation
field operators (13).

\section{Noether's theorem}
We will give Noether's theorem derivation of the linear invariants
(10) (and (14)). Since functions of invariants are also invariants,
the variables
	\begin{eqnarray}
		P_{\vec{k}0}\,(t) & = &\frac { \sqrt{ \omega (\vec{k})
		\ } (A_{\vec{k}} \,(t) - A_{\vec{k}}^{*}\,(t) )} {i
		\sqrt 2},\nonumber \\
		X_{\vec{k}0}\,(t) & = & \frac {A_{\vec{k}}\,(t) +
		A_{\vec{k}}^{*}\,(t) } {\sqrt {2 \omega (\vec{k}) } }
		\end{eqnarray}
are integrals of motion. They may be represented in the matrix form
	\begin{equation}
		\left( \begin{array}{c}
		P_{\vec{k} \, 0}\,(t)\\
		X_{\vec{k} \, 0}\,(t)
		\end{array}\right )
		= \sum _{\vec{k}^\prime}\Lambda _{{\vec k \, \vec
		k}^\prime }\,(t)\left( \begin{array}{c}
		\Pi_{\vec{k}^\prime \,} \\
		\varphi_{\vec{k}^\prime \,}
		\end{array} \right)
		\end{equation}
where the symplectic infinite matrix $\Lambda (t)$ is proportional to
Kronecker term
	\begin{equation}
		\Lambda _{ \vec{k} \, , \vec{k}^{\prime}} \,(t) \sim
		\,\delta _{ \vec{ k} \, , \vec{ k}^\prime} \ .
		\end{equation}
For equal momenta it has the $2\times 2$ form
	\begin{equation}
		\Lambda _{{\vec k \, \vec k}^\prime }\,(t) = \left(
		\begin{array}{clcr}
		\cos (\omega _{\vec k }\,t)&  \omega (\vec{k}) \sin
		(\omega _{\vec k }\,t)\\
		-  \omega^{-1} (\vec{k}) \sin (\omega _{\vec k
		}\,t) & \cos (\omega _{\vec k }\,t) \end{array} \right)
		\ .
		\end{equation}
This expression is the infinite-dimensional generalization of the
invariants for the finite number of degrees of freedom \cite{mmt},
\cite{dod89}, discussed for two-dimensional strings in \cite{dom}.
Now we will derive the invariants following the approach \cite{crm94}.

Now consider an infinitesimal transformation for the field and its
canonically conjugated momentum $ \delta \varphi [ \varphi ( \vec{k} ,
\, t), \, \Pi ( \vec{k} , \, t) , \, \vec{k} , \, t]  , \  \delta \Pi
[ \varphi ( \vec{k} , \, t), \, \Pi ( \vec{k} , \, t) , \, \vec{k} ,
\, t ] $.  This is a symmetry transformation if we have~\cite{cr92}
	\begin{equation}
		\sum_{ \vec{k} } \left\{  \delta \Pi  \ \dot{\varphi}
		( \vec{k} , \, t) + \delta \dot{\varphi} \ \Pi (
		\vec{k} , \, t) - \left[ \delta \varphi \ {\partial H
		\over \partial \varphi ( \vec{k} , \, t) } + \delta
		\Pi \ {\partial H \over \partial \Pi ( \vec{k} , \,
		t) } \right] \right\} = \dot{\Omega} \ .
		\end{equation}
Substituting for the partial derivatives of Hamiltonian (8) we
obtain
	\begin{equation}
		\sum_{ \vec{k} }  \left\{ {d \over dt} \left[ \delta
		\Pi \ \varphi ( \vec{k} , \, t) \right] - \left[
		\delta \dot{\Pi} + \delta \varphi \ \omega^{2} (
		\vec{k}) \right] \ \varphi ( \vec{k} , \, t)
		+ \left[ - \delta \Pi + \delta \dot{\varphi} \right]
		\ \Pi ( \vec{k} , \, t)  \right\} = \dot{\Omega} \ ,
		\end{equation}
where we have integrated by parts the term $ \delta \Pi ( \vec{k} ,
\, t) \ \dot{\varphi} ( \vec{k} , \, t) $ inside the sum over $
\vec{k} $.  Because the fields $ \varphi ( \vec{k} , \, t) $ and
momenta $ \Pi ( \vec{k} , \, t) $ are independent for each value of $
\vec{k} $ we obtain the following conditions
	\begin{eqnarray}
		\Omega & = & \sum_{ \vec{k} } \delta \Pi \ \varphi (
		\vec{k} , \, t) \ , \\
		\delta \dot{\Pi} & = & - \omega^{2} ( \vec{k}) \
		\delta \varphi \ , \\
		\nonumber \\
		\delta \dot{\varphi} & = & \delta \Pi \ ,
		\end{eqnarray}
in order to have a symmetry transformation.  To get the linear
time-dependent invariants we chose the variations as arbitrary
functions of time and the wave vector $\vec{k}$~\cite{crm94}:
	\begin{eqnarray}
		\delta \varphi  [ \varphi ( \vec{k} , \, t), \, \Pi (
		\vec{k} , \, t) , \, \vec{k} , \, t ] &=& {\cal H} (
		\vec{k} , \, t) \ , \\
		\nonumber \\
		\delta \Pi [ \varphi ( \vec{k} , \, t), \, \Pi (
		\vec{k} , \, t) ,
		\, \vec{k} , \, t ] &=& {\cal G} ( \vec{k} , \, t) \ .
		\end{eqnarray}
Substituting these expressions into the equations (22) and (23) we
get a system of coupled equations for the variations in the field and
its conjugated momentum, which it has the same form as the Hamilton
equations.  This system have two independent solutions, and we denote
them by a superscritp in the function, {\it i.e.},
	\begin{eqnarray}
		{\cal H}^{( 1 )} ( \vec{k} , \, t) & = & C ( \vec{k}
		) \ \cos \left(	\omega ( \vec{k} ) \, t \right)  \ ,
		\\
		{\cal H}^{( 2 )} ( \vec{k} , \, t) & = & D ( \vec{k}
		) \ \sin \left( \omega ( \vec{k} ) \, t \right)  \ ,
		\end{eqnarray}
where $C ( \vec{k} )$ and $D ( \vec{k} )$ are arbitrary functions.
The constants of motion \cite{crm94} associated to the symmetry
transformation indicated by (24) and (25) is
	\begin{equation}
		J^{(i)} = \sum_{ \vec{k} } \left\{ {\cal H}^{(i)} (
		\vec{k} , \, t) \ \Pi ( \vec{k} , \, t) - \dot{\cal
		H}^{(i)} ( \vec{k} , \, t) \ \varphi ( \vec{k} , \, t
		) \right\} \ , \qquad i = 1,2 \ .
		\end{equation}
These invariants can be rewritten in matrix form (16), with the
infinite symplectic matrix
	\begin{equation}
		\Lambda_{\vec{k} \ \vec{k}^{\prime} } =
		\delta_{\vec{k} \ \vec{k}^{\prime} } \ \left(
		\begin{array}{clcr}
		{\cal H}^{(1)} ( \vec{k} , \, t ) & - \dot{\cal
		H}^{(1)} ( \vec{k} , \, t ) \\
		{\cal H}^{(2)} ( \vec{k} , \, t ) & - \dot{\cal
		H}^{(2)} ( \vec{k} , \, t ) \end{array} \right) \ .
		\end{equation}
This expression yields infinite number of integrals of motion (10)
and (14) since it is an invariant for arbitrary choices of functions
$D  ( \vec{k} )$ and $ C  ( \vec{k} ) $.  In particular, if the
variations satisfy the initial conditions
	\begin{eqnarray}
		{\cal H}^{(1)}  ( \vec{k} , \, 0 ) & = & 1 \ , \\
		{\cal H}^{(2)}  ( \vec{k} , \, 0 ) & = & - \frac{1}{
		\omega ( \vec{k} ) } \ ,
		\end{eqnarray}
one gets the invariants given in (16).

\section{Conclusions}

We have shown that the linear time-dependent invariants in field
theory, which are useful to construc the propagator and generalized
correlated states, can be obtained through the Noether's theorem.
This approach can be extended to construct linear time-dependent
invariants for scalar fields in non-stationary metrics.

\section*{Acknowledgements}
The authors would like to thank to S. Hojman and M. Man'ko for important
comments on this work.  Also one of us, V. I. M., would like to thank the
Instituto de Ciencias Nucleares, UNAM, for its hospitality.

\end{document}